\documentclass{iaus}
\usepackage{graphicx}

\title[262~~The nature of the LINER in the galaxy NGC 404] %% give here short title %%
{The nature of the LINER in the galaxy NGC 404}

\author[R. B. Menezes, J. E. Steiner, T. V. Ricci \& A. S. Oliveira]   %% give here short author list %%
{Roberto B. Menezes$^1$, Jo\~ao E. Steiner$^1$, Tiago V. Ricci$^1$
 \and Alexandre S. Oliveira$^2$}

\affiliation{$^1$Instituto de Astronomia Geof\'isica e Ci\^encias Atmosf\'ericas, Universidade de S\~ao 
Paulo, Rua do Mat\~ao, 1226,
S\~ao Paulo, SP, Brasil \\[\affilskip]
$^2$IP\&D, Universidade do Vale do Para\'iba, Av. Shishima Hifumi, 2911, S\~ao Jos\'e dos Campos, SP, 
Brasil \\} 

\pubyear{2009}
\volume{262}  %% insert here IAU Symposium No.
\pagerange{1--2}
\jname{Stellar Populations - Planning for the Next Decade}
\editors{A.C. Editor, B.D. Editor \& C.E. Editor, eds.}
\begin{document}

\maketitle

\begin{abstract}
NGC 404, at a distance of 3.4 Mpc, is the nearest S0 galaxy. This galaxy harbors 
a LINER; however, 
since the spectrum does not show a broad H$\alpha$ emission, it is not certain 
that this LINER is a low 
luminosity AGN and its nature is still an open question. HST observations show 
the existence of 
stellar populations with an age of  $3 \times 10^{8}$ years in the galactic 
bulge and with an age of $6-15\times10^{9}$ years 
in the galactic disk. In this work, we present an analysis of the data cube of 
NGC 404 obtained with the 
IFU (Integral Field Unity) of the GMOS (Gemini Multi-Object Spectrograph) on the 
Gemini North telescope. 

\end{abstract}

\firstsection % if your document starts with a section,
\section{Methodology}

We used two methodologies for analyzing the data cube of NGC 404. The first of them was the spectral 
synthesis with the Starlight software (Cid Fernandes et al (2005)). This software performs a spectral 
synthesis to fit an observed spectrum from a set of base spectra. From this spectral synthesis, Starlight 
calculates, among other things, the fraction of light attributed to each stellar population used to 
perform 
the fit. The spectral synthesis was applied to the spectra of each spatial pixel of the data cube.	

The second methodology used for this analysis was the spectral simulations with the Cloudy (developed by 
G. Ferland) 
and Mappings III (Allen et al (2008)) softwares. Cloudy calculates the spectrum emitted by a 
non-equilibrium gas exposed 
or not to a radiation source. Mappings III works in a similar way, but the atomic ionization and 
excitation sources, 
in this case, are shock waves.   

In this work, we tried to reproduce the emission line ratios detected along the field of view of the 
studied object by 
performing simulations with theses softwares.

\section{Results and Discussion}

By using Starlight, we determined the flux and mass fractions corresponding to the stellar populations 
used in the spectral synthesis of the central region of NGC 404. In Figure 1 we can see that most of the 
stellar mass in the central region of this galaxy is due to stars with ages higher than 
$10^{10}$ years and with medium and high metallicities. Therefore, photoionization models involving young 
stars cannot explain the emission spectrum of this galaxy. The graph showing the flux fractions 
corresponding to the stellar populations used in the spectral synthesis appears to suggest the 
existence of several different star formation episodes along the evolution of this galaxy. 

\begin{figure}[h]
% \vspace*{-2.0 cm}
\begin{center}
 \includegraphics[width=5.5in, height=1.6
in]{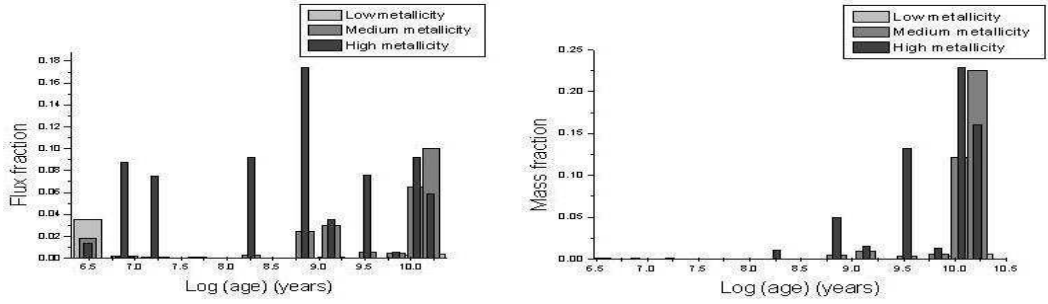} 
% \vspace*{-1.0 cm}
 \caption{(left) Columns graph representing the flux fractions corresponding to the stellar populations 
used in the spectral synthesis. (right) Columns graph representing the mass fractions corresponding to the 
stellar populations used in the spectral synthesis.  }
   \label{Figure 1}
\end{center}
\end{figure}

In all simulations performed with Cloudy for this object, we used an ionizing continuum with a power law 
shape and a spectral 
index of -1.7, a filling factor of 0.01 and an electronic density of $10^{3}$ cm$^{-3}$. In the 
simulations performed with Mappings III, 
we used an electronic density of 1 cm$^{-3}$ and a solar metallicity. 

In Figure 2, we can see that the photoionization by an AGN (with ionization parameters between $10^{-4.5}$ 
and $10^{-3.5}$) can reproduce part of the emission line ratios observed in the data cube of NGC 404. 
However, since there are no observational evidences for the existence of an AGN in this galaxy, a possible 
explication is that there was an AGN in the past, but the accretion stopped and the AGN was extinguished. 
Now, some photons emitted by the AGN are still propagating through the Narrow Line Region, resulting in 
the observed spectrum. The simulations performed with Mappings III show that photoionization models 
involving shock waves (with velocities between 275 and 350 km/s) can reproduce a great part of the 
emission line ratios observed in the data cube of NGC 404 if a low electronic density (1 cm$^{-3}$) is 
assumed. However, this low density seems to be in disagreement with the observed [S II] line ratios.

\begin{figure}[h]
% \vspace*{-2.0 cm}
\begin{center}
 \includegraphics[width=5.5in, height=1.6
in]{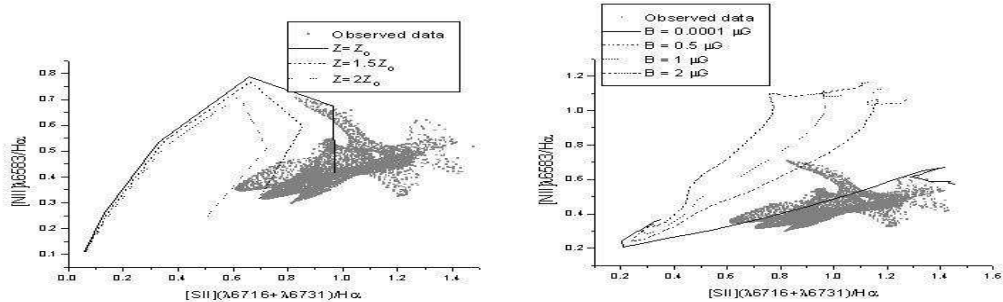} 
% \vspace*{-1.0 cm}
 \caption{(left) Diagnostic diagram [NII]$\lambda$6583/H$\alpha$ x [SII]/H$\alpha$ containing the emission 
line ratios measured from the spectra of the data cube of NGC 404 and the results of the simulations 
performed with Cloudy. The curves were obtained by varying the ionization parameter. (right) Diagnostic 
diagram [NII]$\lambda$6583/H$\alpha$ x [SII]/H$\alpha$ containing the emission line ratios measured from 
the spectra of the data cube of NGC 404. The curves were obtained by varying the shock wave velocities. }
   \label{Figure 2}
\end{center}
\end{figure}


\begin{thebibliography}{}

\bibitem[Allen \etal\ (2008)]{Allen_etal08}
{Allen, M.G., Groves, B.A., Dopita, M.A., Sutherland, R.S. \& Kewley, L.J.}
 2008,
\textit{ApJS}, 178, 20

\bibitem[Cid Fernandes \etal\ (2005)]{CidFernandes_etal05}
{Cid Fernandes, R., Mateus, A., Sodr\'e, L., Stasinska, G.\& Gomes, J.M.}
 2005,
\textit{MNRAS}, 358, 363

\end{thebibliography}
\end{document}